\documentclass{PoS}

\usepackage[english]{babel}
\usepackage{graphicx} 
\usepackage{cite}

\graphicspath{{plots/}} 

\title{Energy dependence of identified hadron spectra and event-by-event fluctuations in p+p interactions from NA61/SHINE at the CERN SPS}

\ShortTitle{Energy dependence of identified hadron spectra and event-by-event fluctuations}

\author{\speaker{Maciej Rybczy\'{n}ski (for the NA61/SHINE Collaboration)}\\
        Institute of Physics, Jan Kochanowski University, PL-25406 Kielce, Poland\\
        E-mail: \email{maciej.rybczynski@ujk.edu.pl}}

\author{The NA61/SHINE Collaboration:\\
N.~Abgrall${}^{22}$,
A.~Aduszkiewicz${}^{23}$,
Y.~Ali${}^{8}$,
T.~Anticic${}^{13}$,
N.~Antoniou${}^{18}$,
J.~Argyriades${}^{22}$,
B.~Baatar${}^{9}$,
A.~Blondel${}^{22}$,
J.~Blumer${}^{5}$,
M.~Bogomilov${}^{4}$
% ...
}

\author{\\
${}^{ 1}$The Universidad Tecnica Federico Santa Maria, Valparaiso, Chile  \\
${}^{ 2}$ETH, Zurich, Switzerland \\
${}^{ 3}$Fachhochschule Frankfurt, Frankfurt, Germany \\
${}^{ 4}$Faculty of Physics, University of Sofia, Sofia, Bulgaria \\
${}^{ 5}$Karlsruhe Institute of Technology, Karlsruhe, Germany \\
${}^{ 6}$Institute for Nuclear Research, Moscow, Russia \\
${}^{ 7}$Institute for Particle and Nuclear Studies, KEK, Tsukuba,  Japan \\
${}^{ 8}$Jagiellonian University, Cracow, Poland  \\
${}^{ 9}$Joint Institute for Nuclear Research, Dubna, Russia \\
${}^{10}$Wigner Research Centre for Physics of the Hungarian Academy of
          Sciences, Budapest, Hungary \\
${}^{11}$LPNHE, University of Paris VI and VII, Paris, France \\
${}^{12}$University of Silesia, Katowice, Poland  \\
${}^{13}$Rudjer Boskovic Institute, Zagreb, Croatia \\
${}^{14}$National Center for Nuclear Research, Warsaw, Poland \\
${}^{15}$St. Petersburg State University, St. Petersburg, Russia \\
${}^{16}$Laboratory of Astroparticle Physics, University Nova Gorica, Nova Gorica, Slovenia  \\
${}^{17}$Jan Kochanowski University in  Kielce, Poland \\
${}^{18}$University of Athens, Athens, Greece \\
${}^{19}$University of Bergen, Bergen, Norway \\
${}^{20}$University of Bern, Bern, Switzerland \\
${}^{21}$University of Frankfurt, Frankfurt, Germany \\
${}^{22}$University of Geneva, Geneva, Switzerland \\
${}^{23}$Faculty of Physics, University of Warsaw, Warsaw, Poland \\
${}^{24}$Warsaw University of Technology, Warsaw, Poland  \\
${}^{25}$University of Wroc{\l}aw, Wroc{\l}aw, Poland  \\
${}^{26}$University of Belgrade, Belgrade, Serbia  \\
${}^{27}$University of California, Irvine, USA  \\ 
}

\abstract{
NA61/SHINE at the CERN SPS is a fixed-target experiment pursuing a rich physics program including measurements for heavy ion, neutrino and cosmic ray physics. The main goal of the ion program is to explore 
the most interesting $T , mu_{B}$ region of the phase diagram of strongly interacting matter. 
We plan to study the properties of the onset of deconfinement and to search 
for the signatures of the critical point. The search is performed by varying collision energy 
(13A-158A GeV/c) and system size (p+p, Be+Be, Ar+Ca, Xe+La). 

Thanks to its large acceptance and excellent particle identification capability NA61/SHINE is well suited 
for performing high-precision particle production measurements as well as for studying event-by-event fluctuations 
in p+p, p+nucleus and nucleus+nucleus collisions.

Preliminary results on p+p interactions at 20, 31, 40, 80 and 158 GeV/c are presented. 
They include inclusive spectra of pi+, pi-, K- and protons as a function of transverse momentum/mass and rapidity 
as well as event-by-event fluctuations of transverse momentum, azimuthal angle and chemical composition.
The new NA61 measurements are compared with the corresponding results of NA49 on central Pb+Pb collisions 
and with predictions of Monte Carlo models.

Finally, the future plans of NA61/SHINE are summarised. 
}

\FullConference{Xth Quark Confinement and the Hadron Spectrum,\\
		October 8-12, 2012\\
		TUM Campus Garching, Munich, Germany}

\begin{document}

\section{\label{sec:intro}Introduction}

NA61/SHINE (\textbf{S}PS \textbf{H}eavy \textbf{I}on and \textbf{N}eutrino \textbf{E}xperiment) is a fixed-target experiment operating since 2007 in the North Area of the CERN Super Proton Synchrotron, using a large-acceptance hadronic spectrometer to study a wide range of phenomena in a number of different hadron+hadron, hadron+nucleus and nucleus+nucleus reactions~\cite{Antoniou:2006mh}. It is the successor of the NA49 experiment, which took data in the years 1994--2002, and reuses most of the NA49 hardware and software~\cite{Afanasev:1999iu}. Large
acceptance (around 50~\% for $p_{T} \le$ 2.5~GeV/c), high momentum resolution ($\sigma(p)/p^{2} \approx 10^{-4}~(\mathrm{GeV/c})^{-1}$) and tracking efficiency (over 95~\%), and excellent particle-identification capabilities ($\sigma(\frac{\mathrm{d}E}{\mathrm{d}x}) / \frac{\mathrm{d}E}{\mathrm{d}x} \approx 4~\%, \sigma(t_{ToF}) \approx 100~\mathrm{ps}$) make it an excellent tool for investigating hadron spectra.

The ultimate goal is to explore the phase diagram of strongly interacting matter in collisions of different systems from p+p, through p+A, to A+A collisions at projectile momenta of 13A, 20A, 30A, 40A, 80A and 158A GeV/c. The
energy scan of p+p reactions was completed in 2009-2011 and preliminary results will be presented.
Data on Be+Be collisions at the three top energies were recorded in 2011, the remaining energies will be taken in 2012. The energy scan of p+Pb, Ar+Ca and Xe+La collisions is foreseen until 2016. In addition a beam energy scan of Pb+Pb collisions is planned. The explored energy range probes an important region in the phase diagram of strongly interacting matter. Indeed, the NA49 collaboration reported signals for the onset of deconfinement encoded in nonmonotonic behavior of excitations functions of several hadronic observables~\cite{Alt:2007aa, Afanasiev:2002mx, Alt:2006jr}. Moreover, the critical point at the location predicted by several theory groups~\cite{Fodor:2004nz} can be probed at SPS energies.

\section{\label{sec:shineDetector}Detector Set-up}
 
\begin{figure}
  \begin{center}
    \includegraphics[width=0.9\columnwidth]{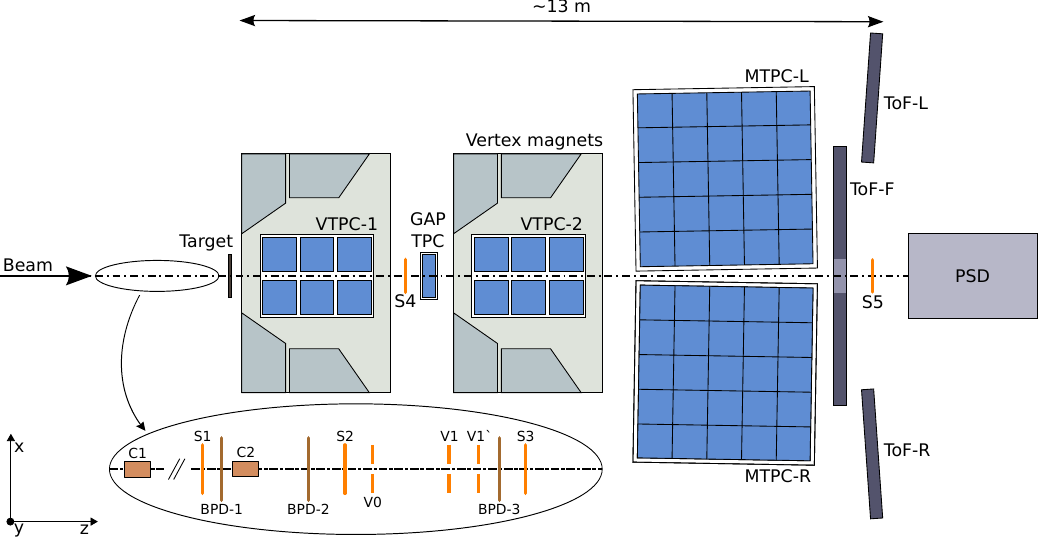}
  \end{center}
  \caption{\label{fig:shineDetector}(Color online) Layout of the NA61/SHINE experimental
  set-up (top view, not to scale).For details see~\cite{Afanasev:1999iu}. }
\end{figure}

Figure~\ref{fig:shineDetector} shows a diagram of the NA61/SHINE apparatus. The primary tracking detectors are four large-volume Time Projection Chambers (TPCs) inherited from NA49. Two of them (Vertex TPCs) are located inside
superconducting magnets along the beam line, the other two (Main TPCs) are placed symmetrically to the beam line further downstream.  Another, smaller chamber (Gap TPC) is placed between the VTPCs. Together, these chambers measure
trajectories and momenta of particles and allow their identification through ionisation energy loss.

Three Time-of-Flight (ToF) walls are positioned just downstream of the MTPCs, two on the sides inherited from NA49 and the central one added by NA61 in 2007. Measurements from these detectors complement particle identification in the momentum range where $\mathrm{d}E / \mathrm{d}x$ alone is ambiguous.

The most downstream detector of the apparatus is the Projectile Spectator Detector (PSD), installed in 2011. The purpose of this calorimeter is to provide precise, high-granularity event-by-event measurements of the energy of
non-interacting fragments of the projectile nuclei, making it possible to determine the centrality of the collision and the orientation of the reaction plane.

Finally, a number of additional detectors installed in the beam line both upstream and downstream of the target monitor beam properties and provide trigger information.

A sophisticated beam line set-up was designed and put in place by the CERN accelerators-and-beams department in collaboration with NA61 for the purpose of providing beryllium beams for SHINE from the fragmentation of Pb ions. At the moment the only primary beams the SPS can provide are protons and lead ions, with argon and xenon to be made possible during the long shutdown of the accelerator complex in 2013-2014. As adding new species of primary beams is an immensely complicated task, beryllium beams for SHINE were instead produced in a fragmentation beam line. Primary Pb ions from the SPS were directed onto a beryllium block far upstream in the beam line and
the resulting fragments of the projectile were guided through a sequence of magnets, collimators, degraders and beam detectors in order to filter out beryllium ions. This configuration was proven to provide highly pure $^7$Be ions over a wide range of Pb-beam energies. 

\section{\label{sec:res}Results}

Data taking for the NA61/SHINE ion program started with
the energy scan of p+p interactions. These data are needed to establish a base
line reference for the results on collisions of light and medium size nuclei. The latter are particularly important
in the search for the critical point of strongly interacting matter.

\subsection{\label{sec:inclSpect}Inclusive particle spectra in p+p interactions}

\begin{figure}
  \begin{center}
    \includegraphics[width=0.49\columnwidth]{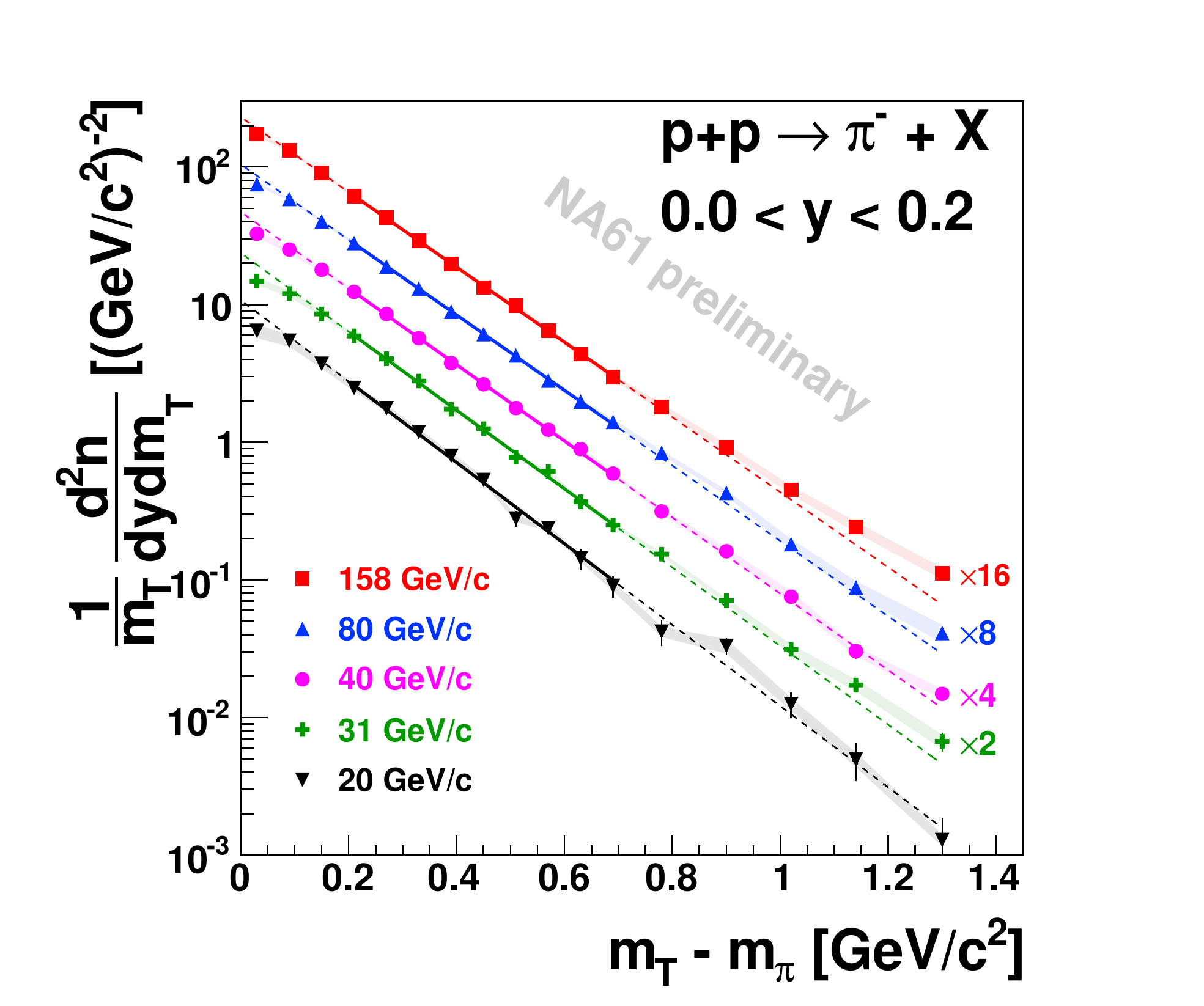}
    \includegraphics[width=0.49\columnwidth]{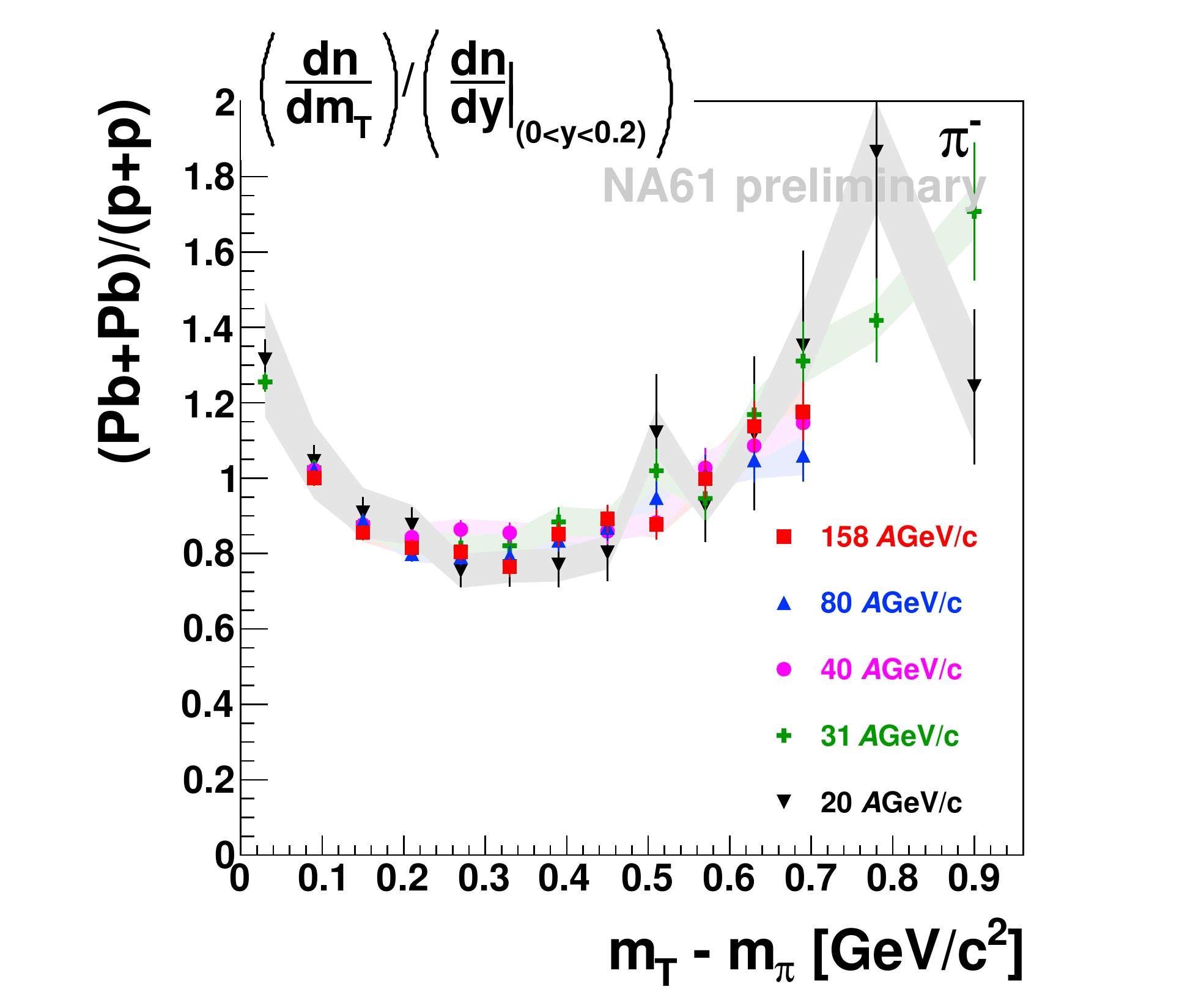}
  \end{center}
  \caption{\label{fig:mtSpect}(Color online) {\it Left}: Transverse-mass spectra at mid-rapidity for negatively charged pions fitted with an exponential function in the $0.2 < m_T-m_\pi < 0.7~{\rm GeV/c^2}$ interval. {\it Right}: The ratio of Pb+Pb to p+p transverse mass spectra. Results refer to inelastic p+p interactions at $20-158~{\rm GeV/c}$.}
\end{figure}

\begin{figure}
  \begin{center}
    \includegraphics[width=0.49\columnwidth]{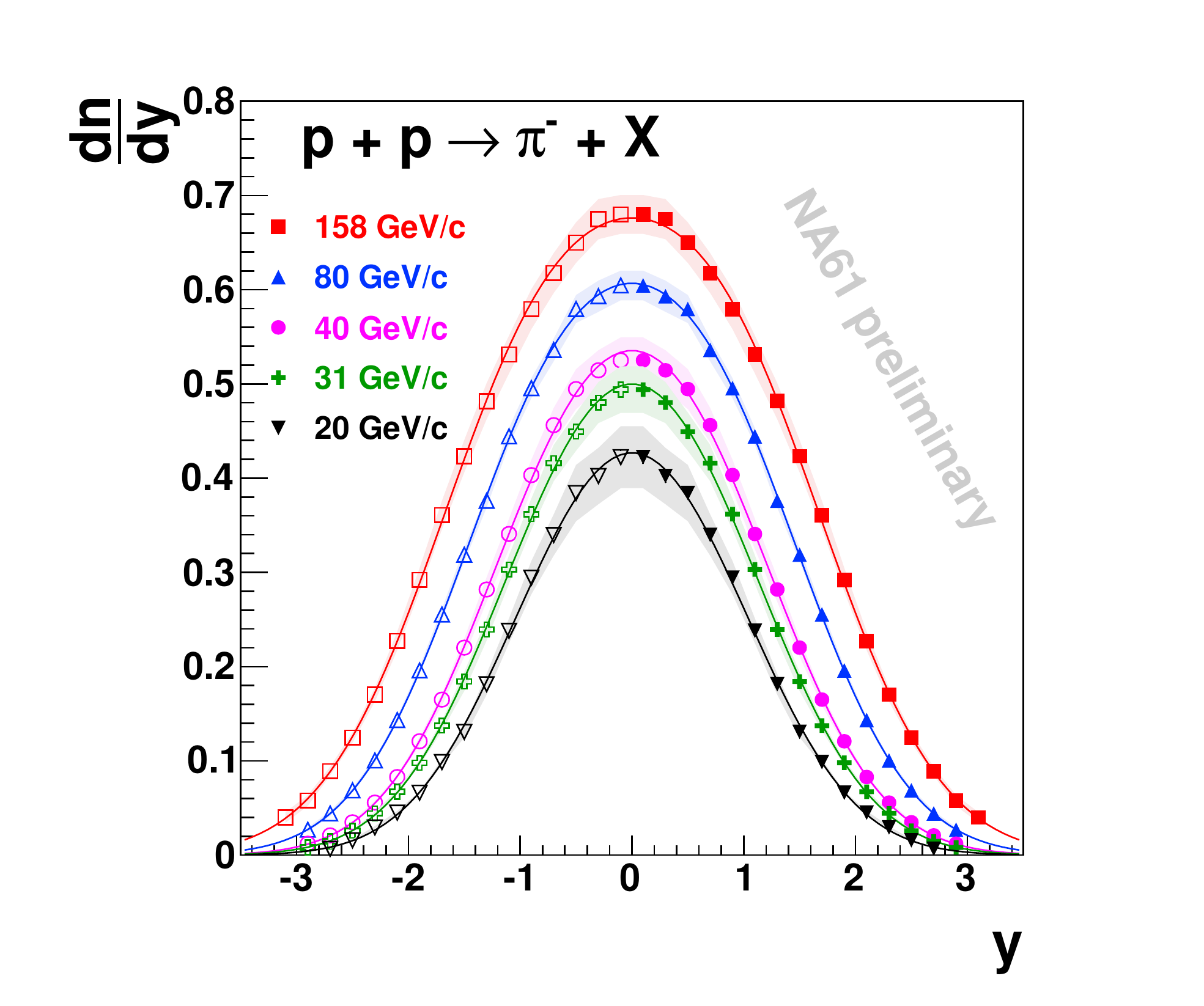}
    \includegraphics[width=0.49\columnwidth]{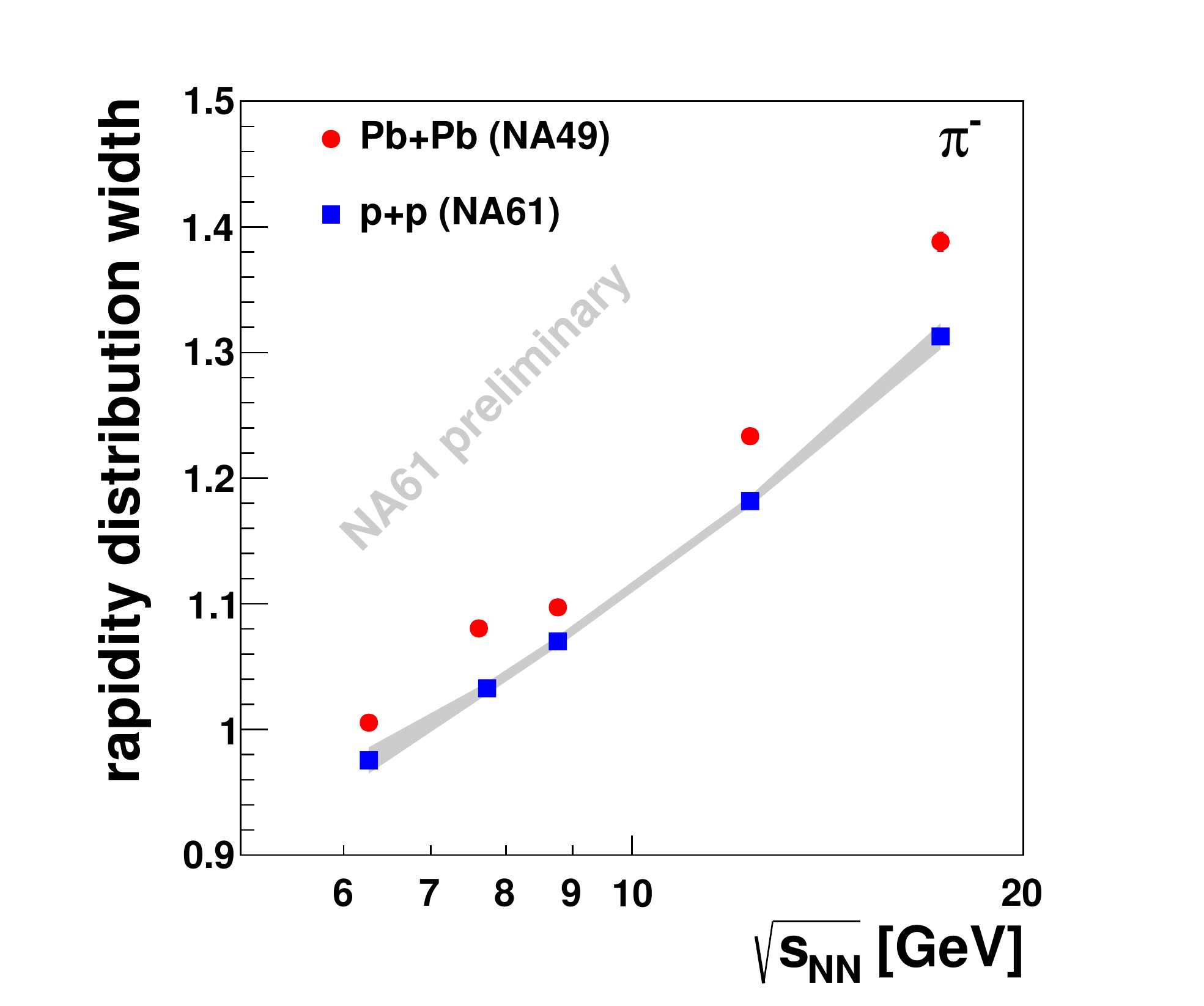}
  \end{center}
  \caption{\label{fig:rapSpect}(Color online) {\it Left}: Rapidity spectra of negatively charged pions fitted
with the sum of two Gaussian functions. {\it Right}: The width of rapidity spectra of $\pi^-$ produced in p+p interactions at $20-158~{\rm GeV/c}$ compared to NA49 results from central Pb+Pb collisions~\cite{Afanasiev:2002mx, Alt:2007aa}.}
\end{figure}

First measurements of $\pi^-$ spectra in p+p collisions at 20, 31, 40, 80, and 158 GeV/c are available. Figure ~\ref{fig:mtSpect} (left) presents the transverse mass ($m_T$) spectra at mid-rapidity 
for negatively charged pions in inelastic p+p collisions. The results are corrected for reconstruction inefficiencies, acceptance, and feed-down using Monte Carlo simulations. Figure~\ref{fig:mtSpect} (right) shows 
the ratios of $m_T$ spectra from central (7\%) Pb+Pb to inelastic p+p collisions~\cite{Afanasiev:2002mx, Alt:2007aa}. The observed differences do not change with collision energy and could be due to transverse collective flow present in Pb+Pb. The missing part of the $m_T$ spectra are extrapolated by fitting the distributions with an exponential function. This allows to obtain the $m_T$ integrated rapidity spectra depicted in Figure~\ref{fig:rapSpect} (left). The spectra are well described by a sum of two symmetrically displaced Gaussian functions. When compared to Pb+Pb data the widths of $\pi^-$ rapidity distributions for p+p are clearly narrower, see Figure~\ref{fig:rapSpect} (right). The possible reasons are isospin effects and/or longitudinal collective flow in central Pb+Pb collisions. The two-Gaussian fit allows to extrapolate the rapidity spectrum and calculate the mean $\pi^-$ multiplicity.

\begin{figure}
  \begin{center}
    \includegraphics[width=0.48\columnwidth]{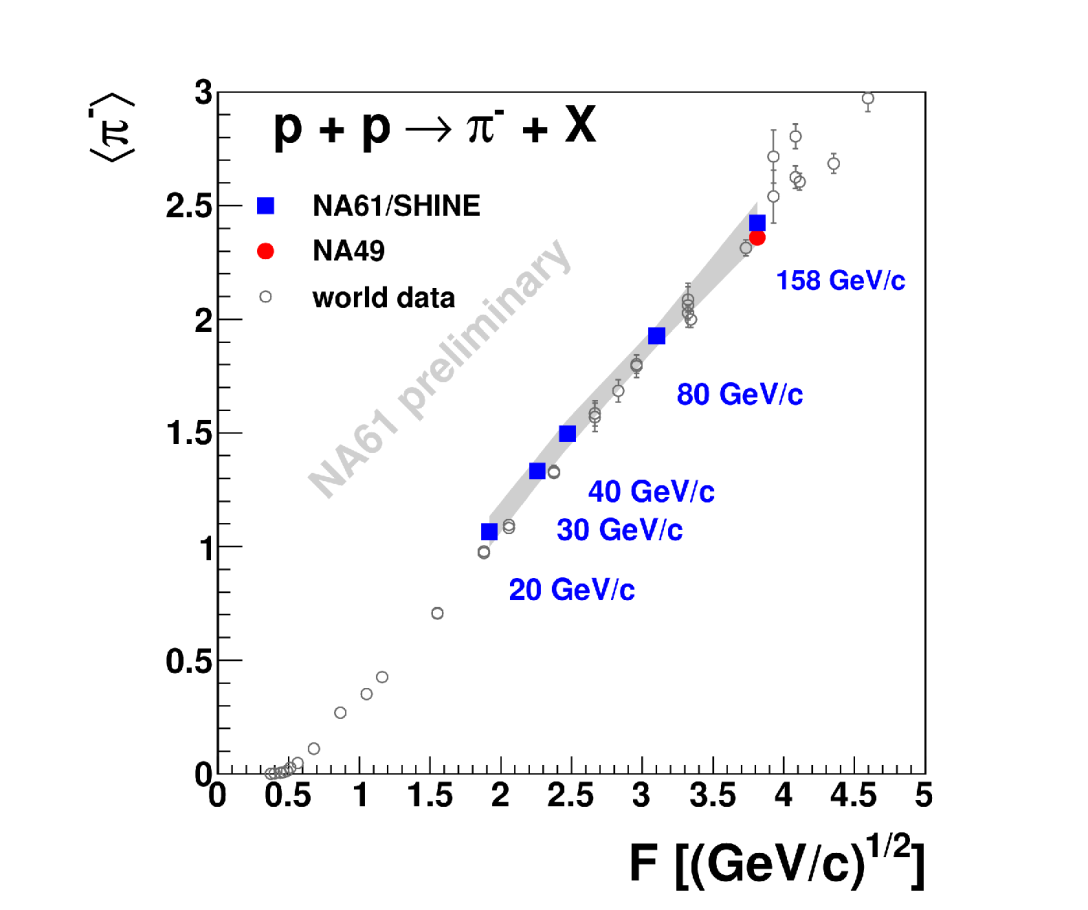}
    \includegraphics[width=0.5\columnwidth]{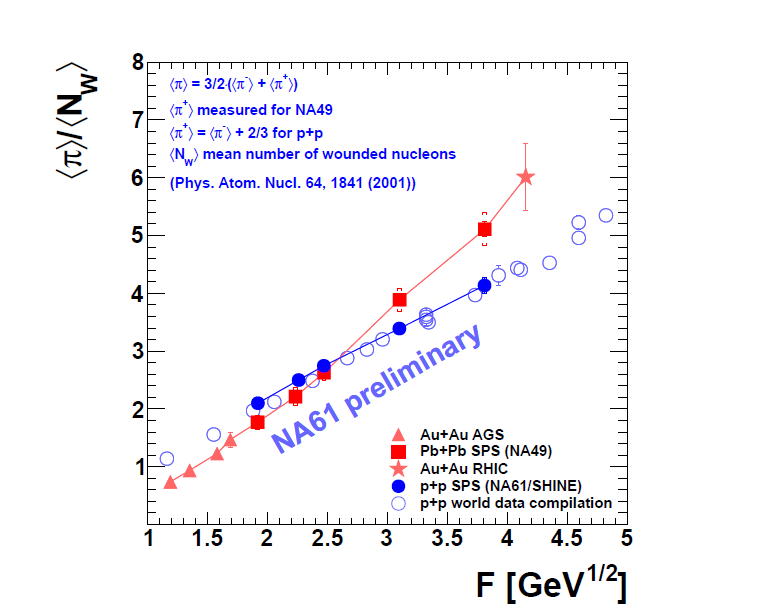}
  \end{center}
  \caption{\label{fig:mult}(Color online) {\it Left}: Mean negatively charged pion multiplicity. {\it Right}: Mean pion multiplicity per participant nucleon. Both are shown as function of the Fermi variable $F \approx s^{0.25}$.}
\end{figure}

Figure~\ref{fig:mult} (left) shows that the NA61/SHINE pion multiplicities are in good agreement with the world data.
Figure~\ref{fig:mult} (right) presents the "kink" plot, showing the total pion multiplicity normalized to the number of wounded nucleons versus the Fermi variable $F \approx s^{0.25}$. Unlike for p+p interactions, a change of slope is visible around 30A GeV in central Pb+Pb and Au+Au collisions. Such an increase can be explained by an
increase of the effective number of degrees of freedom when going from hadron gas to QGP as a
consequence of the activation of partonic degrees of freedom. Also, preliminary results on $p_T$ and rapidity
spectra of charged pions, negatively charged kaons, and protons were obtained in inelastic
p+p interactions at 40, 80 and 158 GeV/c. These will provide the baseline for the study of the properties of the onset of
deconfinement by looking for the kink, horn, step and dale structures~\cite{Gazdzicki:1998vd,Anticic:2009wd,Gazdzicki:2010iv} in collisions of light and
intermediate mass nuclei (Be+Be, Ar+Ca, Xe+La).

\subsection{\label{sec:ebeFluct}Event by Event fluctuations in Pb+Pb and p+p collisions}

The value of $\mathrm{d}E / \mathrm{d}x$ does not allow to identify each particle uniquely as the $\mathrm{d}E / \mathrm{d}x$ distributions overlap. The identity method~\cite{Rustamov:2012bx} was developed to extract second and third moments (pure and mixed) of identified particle multiplicity distribution corrected for this imperfect identification. First fluctuation measurements in p+p give a unique opportunity to compare Pb+Pb and p+p results at the SPS energies.

Preliminary results on multiplicity fluctuations of $\pi = \pi^+ + \pi^-$, $K = K^+ + K^-$, and $p = p + \bar{p}$ were obtained for p+p interactions at 31, 40, 80 and 158 GeV/c. From the first and second corrected moments of the multiplicity distributions, $N_i$ and $N_i^2$, ($i=\pi, K, p$), the scaled variance 

\begin{equation}
\omega_i=\frac{\langle N_i^2\rangle - \langle N_i\rangle^2}{\langle N_i\rangle }
\end{equation}

was computed. For Poisson multiplicity distributions $\omega_i = 1$ and does not depend on the number of wounded nucleons. However, it is sensitive to the fluctuations of the wounded nucleon number. This may distort the comparison of p+p data
with results from nucleus+nucleus collisions.

\begin{figure}[!h]
  \begin{minipage}[t]{7.5cm}
    \begin{center}
      \centerline{\includegraphics[width=0.95\columnwidth]{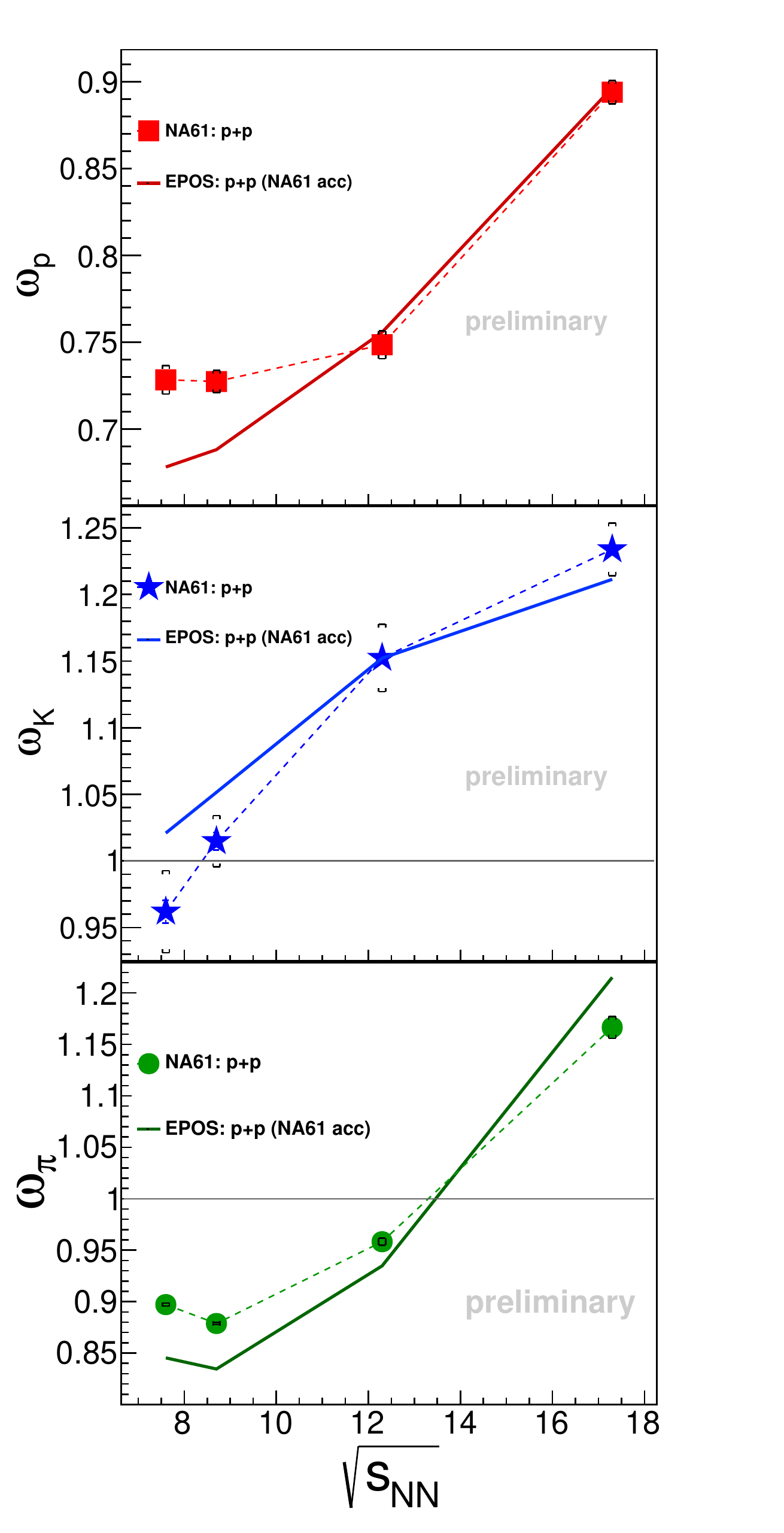}}
      \nopagebreak[4]
      \vspace{-5mm}
      \caption{(Color online) Scaled variance for $p+\bar{p}$ (upper panel), $K^+ + K^-$ (middle panel) and  $\pi^+ + \pi^-$ (bottom panel) multiplicity distributions for p+p
collisions as a function of center of mass energy. The NA61/SHINE results are shown by filled symbols connected by dashed lines and compared to EPOS model calculations depicted by a solid lines.}
      \label{fig:omega}
    \end{center}
  \end{minipage}
  \hspace{0.5cm}
\begin{minipage}[t]{7.5cm}
    \begin{center}
      \centerline{\includegraphics[width=0.99\columnwidth]{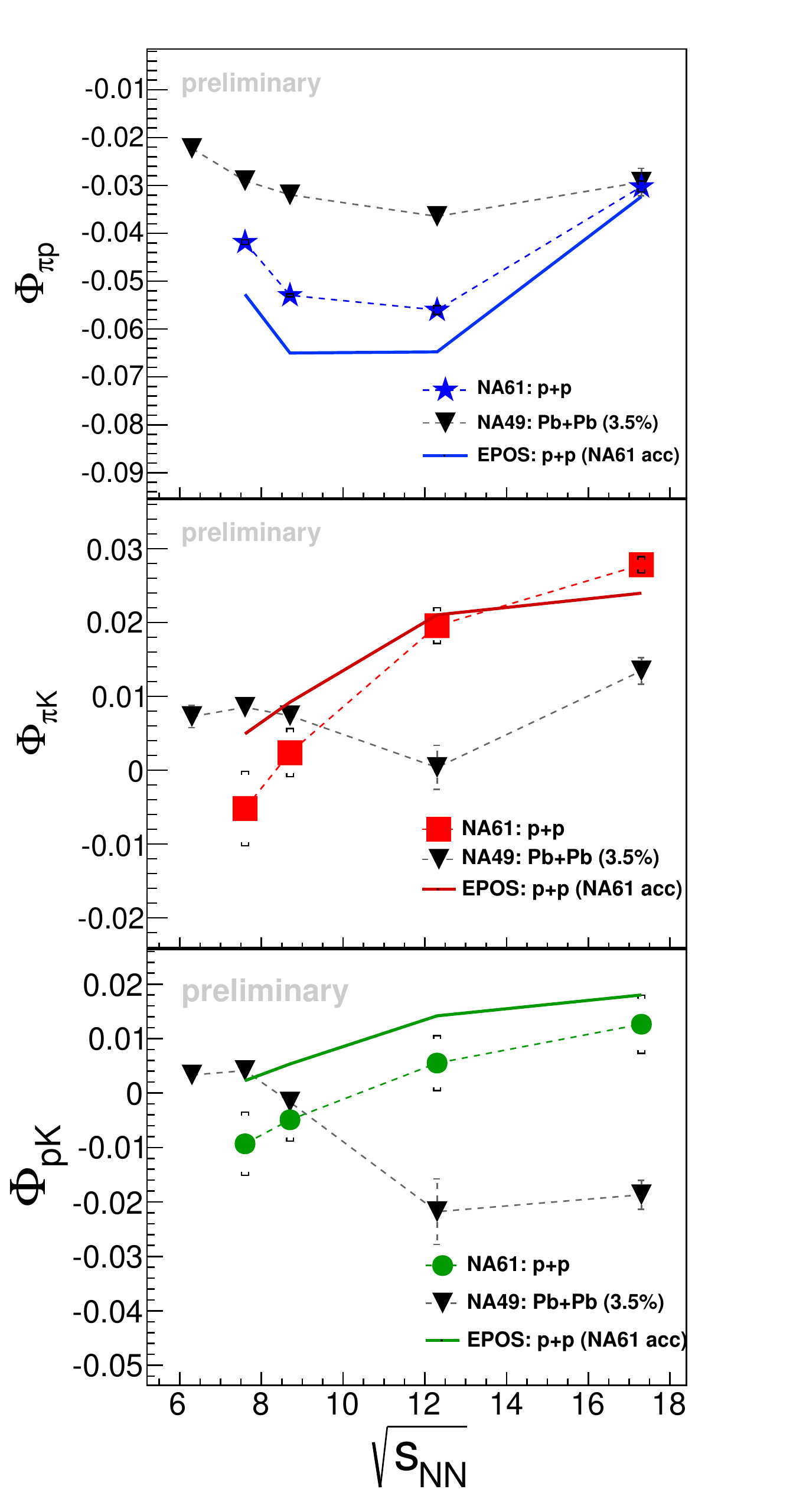}}
      \nopagebreak[4]
      \vspace{-5mm}
      \caption{(Color online) $\Phi_{ij}$ measure as a function of center of mass energy for p+p (NA61/SHINE) and central Pb+Pb (NA49) collisions. The p+p data are compared with the EPOS model predictions (solid lines).}
      \label{fig:phi}
     \end{center}
  \end{minipage}
\end{figure}

All studied scaled variances increase with increasing collision energy (Figure~\ref{fig:omega}). For kaons $\omega > 1$ for higher energies in agreement with predictions of the EPOS model. This may be caused by the correlation in $K^+$ and $K^-$ production due to strangeness conservation. For protons $\omega$ is below 1, probably related to baryon number conservation. A comparison of NA61/SHINE results on multiplicity fluctuations (i.e. $\omega_{\pi}$ ) with the magnitude of fluctuations expected at the critical point~\cite{Grebieszkow:2009jr} suggests that the systematic and statistical errors are small enough for a sensitive search of the critical point in NA61/SHINE.



In order to compare results for p+p and central Pb+Pb collisions, the strongly intensive measure $\Phi_{ij}$~\cite{Gorenstein:2011vq, Gazdzicki:1997gm} defined for two hadron types, $i$ and $j$, was chosen. It is defined as:

\begin{equation}
\Phi_{ij}=\frac{\sqrt{\langle N_i \rangle \langle N_j \rangle}}{\langle N_i + N_j \rangle} \cdot 
\left( \sqrt{\Sigma^{ij}}-1 \right)
\end{equation}
where $\Sigma^{ij}=\langle N_i \rangle \omega_j + \langle N_j \rangle \omega_i + 2 \cdot 
\left(\langle N_{ij} \rangle - \langle N_i \rangle \langle N_j \rangle \right) \cdot \langle N_i + N_j \rangle$. 
As a strongly intensive measure, $\Phi_{ij}$ is not only independent of number of wounded nucleons or volume but also of their fluctuations. Figure~\ref{fig:phi} shows the energy dependence of $\Phi_{ij}$ for combinations of two hadron types: $\pi p$, $\pi K$, and $p K$. When no inter-particle correlations are present $\Phi_{ij} = 0$. For $\pi K$ and $p K$ the values of $\Phi_{ij}$ increase with increasing energy. There is a minimum for $\Phi_{\pi p}$ between 7.3 and 8.7 GeV. A similar but weaker effect is visible in Pb+Pb interactions. It also appears in the EPOS model. The increase
of $\Phi_{\pi K}$ for p+p interactions is not visible in Pb+Pb collisions. $\Phi_{p K}$ shows a clear difference between results for p+p, which increase with increasing energy, and for Pb+Pb which decrease with increasing energy. Both dependences cross zero at the same energy $\sqrt{s_{NN}}\approx 8.7$~GeV.

\section{\label{sec:Summ}Summary}
We presented preliminary results from NA61/SHINE on particle production in inelastic p+p collisions. These data represent the
first phase of the scan of the phase diagram of strongly interacting matter. Inclusive particle spectra and mean
multiplicities were obtained. Event-by-event chemical fluctuations were analyzed using a novel approach, the Identity Method. 
These results will serve as an important reference for A+A collisions.

\vspace{0.5cm}

This work was supported by the Polish Ministry of Science and Higher Education under the grant 667/N-CERN/2010/0 and the Foundation for Polish Science - MPD program co-financed by the European Union within the European Regional Development Fund.


\begin{thebibliography}{99}

\bibitem{Antoniou:2006mh}
N.~Antoniou {\it et al.} [NA61/SHINE Collaboration], CERN-SPSC-2006-034,
  CERN-SPSC-P-330  (2006) 

\bibitem{Afanasev:1999iu}
S.~Afanasev {\it et al.} [NA49 Collaboration, Nucl.Instrum.Meth. A {\bf 430}, 210 (1999) 

\bibitem{Afanasiev:2002mx} 
  S.~V.~Afanasiev {\it et al.}  [NA49 Collaboration],
  Phys.\ Rev.\ C {\bf 66}, 054902 (2002)
  [nucl-ex/0205002].

\bibitem{Alt:2007aa} 
  C.~Alt {\it et al.}  [NA49 Collaboration],
  Phys.\ Rev.\ C {\bf 77}, 024903 (2008)
  [arXiv:0710.0118 [nucl-ex]].

\bibitem{Alt:2006jr} 
  C.~Alt {\it et al.}  [NA49 Collaboration],
  Phys.\ Rev.\ C {\bf 75}, 064904 (2007)
  [nucl-ex/0612010].

\bibitem{Fodor:2004nz} 
  Z.~Fodor and S.~D.~Katz,
  JHEP {\bf 0404}, 050 (2004)
  [hep-lat/0402006].

\bibitem{Gazdzicki:1998vd} 
  M.~Gazdzicki and M.~I.~Gorenstein,
  Acta Phys.\ Polon.\ B {\bf 30}, 2705 (1999)
  [hep-ph/9803462].

\bibitem{Anticic:2009wd} 
  T.~Anticic {\it et al.}  [NA49 Collaboration],
  Eur.\ Phys.\ J.\ C {\bf 65}, 9 (2010)
  [arXiv:0904.2708 [hep-ex]].

\bibitem{Gazdzicki:2010iv} 
  M.~Gazdzicki, M.~Gorenstein and P.~Seyboth,
  Acta Phys.\ Polon.\ B {\bf 42}, 307 (2011)
  [arXiv:1006.1765 [hep-ph]].

\bibitem{Rustamov:2012bx} 
  A.~Rustamov and M.~I.~Gorenstein,
  Phys.\ Rev.\ C {\bf 86}, 044906 (2012)
  [arXiv:1204.6632 [nucl-th]].

\bibitem{Grebieszkow:2009jr} 
  K.~Grebieszkow [NA49 Collaboration],
  Nucl.\ Phys.\ A {\bf 830}, 547C (2009)
  [arXiv:0907.4101 [nucl-ex]].

\bibitem{Gorenstein:2011vq} 
  M.~I.~Gorenstein and M.~Gazdzicki,
  Phys.\ Rev.\ C {\bf 84}, 014904 (2011)
  [arXiv:1101.4865 [nucl-th]].

\bibitem{Gazdzicki:1997gm} 
  M.~Gazdzicki,
  Eur.\ Phys.\ J.\ C {\bf 8}, 131 (1999)
  [nucl-th/9712050].

\end{thebibliography}
\end{document}